\documentclass[english]{aa1}
\usepackage{times}
\usepackage[T1]{fontenc}
\usepackage[latin1]{inputenc}
\usepackage{amsmath}
\usepackage{graphicx}
\usepackage{amssymb}

\makeatletter

\providecommand{\tabularnewline}{\\}

\usepackage{multirow}

\usepackage{slashbox}

\usepackage{babel}
\makeatother
\begin{document}
\authorrunning{J. Cant\'o, S. Curiel and E. Mart\'inez-G\'omez}

\titlerunning{A simple algorithm for optimization and model fitting}

\title{A simple algorithm for optimization and model fitting: AGA (Asexual
Genetic Algorithm)}

\author{J. Cant\'o, S. Curiel \and E. Mart\'inez-G\'omez}

\offprints{E. Mart\'inez-G\'omez}

\institute{Instituto de Astronom\'ia, Universidad Nacional Aut\'onoma de M\'exico,
Apdo. Postal 70-264, Ciudad Universitaria, Coyoac\'an, 04510, Mexico
\\
\email{scuriel@astroscu.unam.mx; affabeca@gmail.com} \\
}

\date{Received ; Accepted}

\maketitle

\section{Introduction}

Mathematical optimization can be used as a computational tool in deriving
the optimal solution for a given problem in a systematic and efficient
way. The need to search for parameters that cause a function to be
extremal occurs in many kinds of optimization. The optimization techniques
fall in two groups: \emph{deterministic} (Horst and Tuy, \cite{Horst})
and \emph{stochastic} (Guus, Boender and Romeijn ,\cite{Guus}). In
the first group, we have the classical methods that are useful in
finding the optimum solution or unconstrained maxima or minima of
continuous and twice-differentiable functions. In this case, the optimization
consists of identifying points where the gradient of the objective
function is zero and using the Hessian matrix to classify the type
of each point. For instance, if the Hessian matrix is positive definite,
the point is a local minimum, if it is negative, the point is a local
maximum, and if if indefinite, the point is some kind of saddle point.
However, the classical methods have limited scope in practical applications
since some involve objective functions that are not continuous and/or
not differentiable. For these reasons, it is necessary to develop
more advanced techniques that belong to the second group. Stochastic
models rely on probabilistic approaches and have only weak theoretical
guarantees of convergence to the global solution. Some of the most
useful stochastic optimization techniques include: adaptive random
search (Brooks, \cite{Brooks}), clustering methods (Törn, \cite{Torn}),
evolutionary computation that includes genetic algorithms, evolutionary
strategies and evolutionary programming (Fogel, Owens and Walsh, \cite{Fogel};
Schwefel, \cite{Schwefel}; Goldberg, \cite{Goldberg}; McCall, \cite{McCall}),
simulated and quantum annealing (Kirkpatrick, Gellatt and Vecchi,
\cite{Kirkpatrick}), and neural networks (Bounds, \cite{Bounds}).

We present a simple algorithm for optimization (finding the values
of the variables that maximize a function) and model fitting (finding
the values of the model parameters that fit a set of data most closely).
The algorithm may be called genetic, although it differs from standard
genetic algorithms (Holland, \cite{Holland}) in the way that new
generations are constructed. Standard genetic algorithms involve \emph{sexual
reproduction}, that is, the reproduction by the union of male and
female reproductive individuals. Instead, our algorithm uses \emph{asexual
reproduction}, in which offspring are produced by a single parent
(as in the fission of bacterial cells).

The paper is organized as follows. In Sect. 2, we present and describe
the main characteristics of our algorithm called AGA (\textbf{A}sexual
\textbf{G}enetic \textbf{A}lgorithm). In Sect. 3, we apply the algorithm
to two kinds of problems: maximization of complicated mathematical
functions and a model fitting procedure. In the latter group, we consider
two examples taken from astronomy: a) the orbital fitting of exoplanets,
and b) the model fitting of the Spectral Energy Distribution (SED)
observed in a Young Stellar Object (YSO). In both cases, we minimize
their corresponding chi-square function. In Sect. 4, we summarize
and discuss the results for each case.

\section{Description of the AGA (Asexual Genetic Algorithm)}

We consider the problem of finding the absolute maxima of a real function
of \emph{N} variables i.e., identifying the values of the \emph{N}
variables (the coordinates of a point in the space of \emph{N} dimensions)
for which the function attains its maximum value. It is assumed that
the absolute maximum is inside a bounded region \emph{V} where the
function is defined.

\begin{figure}
\centering

\includegraphics[width=10cm,height=10cm,keepaspectratio]{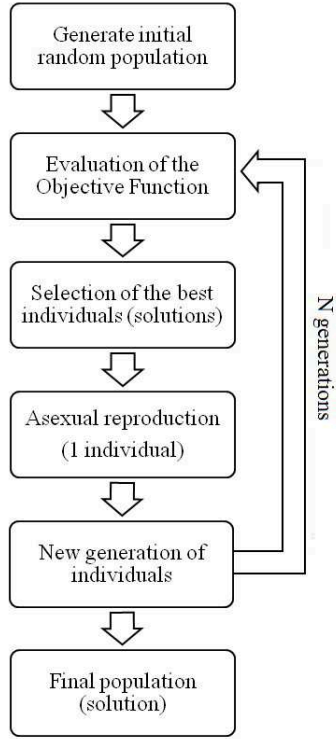}

\caption{Basic diagram for the implementation of the Asexual Genetic Algorithm
(AGA). First we generate a random initial population. Then, we evaluate
the fitness of each individual in this population and select those
which have the highest fitness. A new generation is constructed by
an asexual reproduction (see text) which replaces the older one. If
the stopping criteria are met, we stop. If not, we use these individuals
as an initial population and start again.}

\label{fig:1}
\end{figure}

Our algorithm proceeds in the following way (see also Fig. \ref{fig:1}):

\begin{enumerate}
\item Construct a random initial population. The initial population is a
set of $N_{0}$ randomly generated points (in the context of evolutionary
algorithms, they are also called individuals) within the region \emph{V}.
\item Calculate the fitness of each individual in the population. The fitness
is calculated by evaluating the function at each point.
\item Select a subset of individuals with the highest fitness. Rank the
points according to the value of the function, and choose a subset
of $N_{1}$ points with the highest values of the function.
\item Construct a new population using the individuals in the subset. Generate
$N_{2}$ random points within a previously selected vicinity \emph{E}
around each of the selected points.
\item Replace the source population with the new population. The new population
is the set of $\left(N_{1}\times N_{2}\right)+N_{1}$ points that
results from step 4 plus a clone of each parent. We may choose (as
we did) $N_{1}$ and $N_{2}$ such that $\left(N_{1}\times N_{2}\right)+N_{1}=N_{0}$,
keeping the size of the population $N_{0}$ unchanged for each generation.
In this way, one can devise an iterative procedure.
\item If the stopping criteria (accuracy, maximum number of generations,
etc.) have not been met, return to step 2.
\end{enumerate}
It is clear that \textbf{}this \textbf{}presented algorithm in many
aspects resembles standard genetic algorithms. The key difference
is the way the new population is constructed: in this version, we
propose an \emph{asexual reproduction with mutation} in the sense
that the offspring of a parent is a point randomly selected within
a narrow neighborhood of the (single) parent point. Hence, the name
for our algorithm is AGA, meaning \textbf{A}sexual \textbf{G}enetic
\textbf{A}lgorithm (Fig. \ref{fig:1}). We note that in the new generation
we always include a clone of the parent, that is, we always keep the
original seed points to be used for the next generation whenever the
parents are more well suited than their offspring.

For the vicinity \emph{E} of each selected point, we used rectangular
(hyper)boxes of decreasing size. The box around each point was centered
on the point and had a side length $2\Delta_{ij}$ along direction
\emph{i} in the generation \emph{j}. In particular, we take,

\begin{equation}
\Delta_{i\, j}=\Delta{}_{i\,0}p^{j},\label{eq:1}\end{equation}

\noindent where $2\Delta_{i0}$ is the initial length of the box along
direction \emph{i,} and \emph{p} is a fixed numerical value less than
unity (which can be called the {}``convergency factor''). In this
way, the length of each box side decreases by a factor \emph{p} in
each generation. For instance, if we want the side length of the box
to decrease by a factor 2 after 10 generations then,

\begin{equation}
p=\left(\frac{1}{2}\right)^{\frac{1}{10}}=0.9330\label{eq:2}\end{equation}

Decreasing the size of the vicinity \emph{E} each generation is intended
to achieve the highest possible accuracy for the position of the point
at which the function attains its absolute maxima. The speed with
which the AGA finds a solution depends, of course, on the factor \textit{p};
the lower the value of \textit{p,} the faster the solution found.
However, if \textit{p} is too low, the sampling area around the points
may decrease so fast that the AGA has no time to migrate to the true
solution. On the other hand, if \textit{p} is too high, many of the
offspring will never reach the solution within the convergency criterion.
As a consequence of this, the error in the solution will be high and
in some cases, the solution will not be reached. The adequate value
for \textit{p} depends on the problem itself. The optimization of
\textit{p} can be achieved by trial and error. However, we have found
that a value between 0.4 and 0.6 is adequate for all the tested problems.

An alternative way of choosing the box size consists of employing
the standard deviation of the points contained in the subset $N_{1}$
along each dimension. In such a case, the length of the sides of the
box naturally decreases as the algorithm converges. This method works
quite efficiently for problems with a few dimensions. Interestingly,
the length of the box size in this case decreases following a power
law such as that in Eq. \ref{eq:1}, with a moderately high values
of \textit{p} (between 0.5-0.8) for the first few generations, abruptly
changing to a much lower values (0.2-0.3) for the rest of the generations.

In the case of problems with the large number of parameters to be
estimated, we added an iterative method to the scheme presented above.
This iterative method consisted of performing a series of runs (each
one following the scheme shown in Fig. \ref{fig:1}) in such a way
that the resultant parameters of a run are taken as the initial {}``guess''
for the next run. This procedure may be equivalent to performing a
single run allowing for additional generations, but this is not the
case. The key difference is that for each run in the iterative procedure,
the sizes of the sampling boxes are reset to their initial values,
i. e., each run starts searching for solution using boxes of the same
size as those used in the first run but centered on improved initial
values. We consider to have reached the optimal solution when the
values of the parameters do not change considerably (within a tolerance
limit) after several iterations, and the $\chi^{2}$ is found to have
reached a limiting value (see sect. 3.2.1 and Fig. \ref{fig:7} for
an example).

We find that this iterative strategy guarantees the convergence to
the optimal solution, since it avoids the potential danger of using
values of \textit{p} that do not allow the AGA to drift (migrate)
to the {}``true'' solution within a single run. Problems that involve
the finding of a large number of parameters are potentially subject
to this risk, since each parameter may require a different value of
\textbf{}\textit{p.} Furthermore, problems with a large number of
parameters become particularly difficult when the values of the different
parameters differ by several orders of magnitude, as in the fitting
of the orbits of exoplanets and the fitting of the SED of YSOs (see
Examples 1 and 2 in Sect. 3). In this case, the iterative method has
proven to be particularly useful; the solution usually improves considerably
after several iterations.

In the following section, we describe some applications of AGA. We
have divided the applications into two groups depending on the type
of problem to solve: the maximization of complicated functions and
model fitting in astronomy.

\section{Applications of the AGA}

We separate the optimization problems in two groups. In the first
group, we consider functions of two variables where many classical
optimization methods present formidable difficulties in finding the
global maximum. In the second group, we show two typical examples
of model fitting in astronomy that can be treated as minimization
procedures.

\subsection{Optimization of functions of two variables}

There are many examples of functions that are not easy to optimize
with classical techniques such as the simplex method, the gradient
(or Newton-Raphson) method, the steepest ascent method (Everitt, \cite{Everitt}),
among others. In such cases, the existence of many maxima (or minima)
and the sharpness of the peaks can represent a serious problem. Because
of this, the standard Genetic Algorithms or GA are successfully applied
in searching for the optimal solution. We consider some typical examples
treated by this technique, which are shown below.

\subsubsection{Example 1}

We consider the following function (Charbonneau, \cite{Charbonneau}):

\begin{equation}
f\left(x,y\right)=\left[16x\left(1-x\right)y\left(1-y\right)sin\left(n\pi x\right)sin\left(n\pi y\right)\right]^{2}\label{eq:3}\end{equation}

\noindent where \textbf{}the variables $x,\, y\in\left[0,1\right]$
and $n\in\mathbb{N}-\left\{ 0\right\} $. Identifying the global maximum
of this function for large n is a difficult task because there are
many local maxima that differ little in value but are separated by
deep {}``valleys'' in the two-dimensional landscape. Techniques
such as the steepest ascent/descent and the conjugate gradient method
are local methods that work well if $f\left(x,\, y\right)$ is a smooth
function that can be differentiated at least once and a single maximum
exists in the domain under consideration.

\begin{figure*}
\centering

\includegraphics[width=12cm,height=12cm,keepaspectratio]{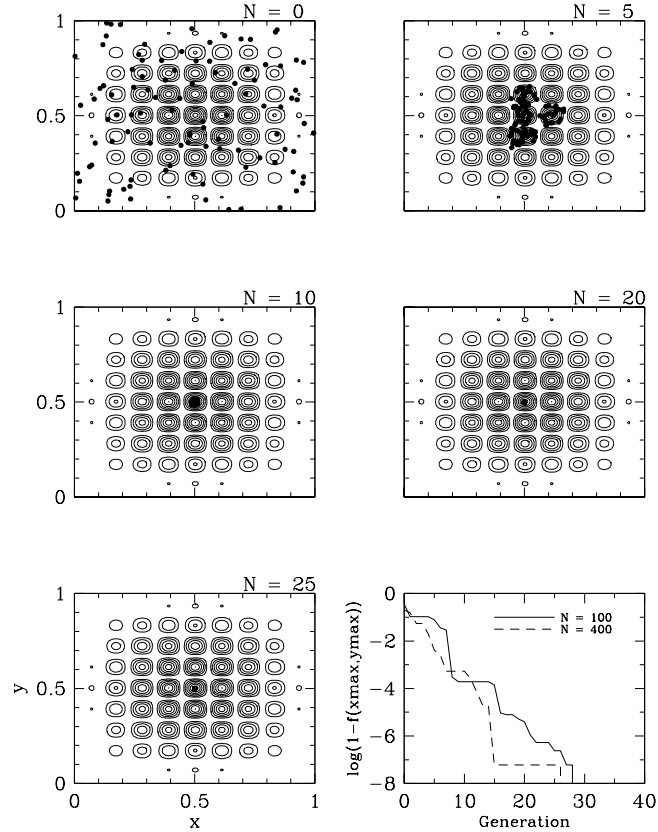}

\caption{A solution to the model optimization problem based on the idea of
the Asexual Genetic Algorithm. The first five panels show the elevation
contours of constant \emph{f} (Eq. \ref{eq:3} with $n=9$) and the
population distribution of candidate solutions (each one contains
100 points), starting with the initial random population (in the standard
GA it is defined as the \emph{{}``genotype''}) in (a) and proceeding
on through the 25th generation on (e). In the sixth panel, we show
the evolution of the fittest \emph{{}``phenotype''} assuming \emph{}two
sizes of the population, $N_{0}=100$ and $N_{0}=400$. }

\label{fig:2}
\end{figure*}

In Fig. \ref{fig:2}, we show the solution to this problem ($n=9$)
by applying AGA after fixing the initial population size, number of
parents, number of descendants, and convergency factor (Table \ref{tab:1}).
Our graphs have the same format as those presented by Charbonneau
(\cite{Charbonneau}) to facilitate their comparison.

\begin{table}
\centering

\caption{\label{tab:1} Values of the initial parameters used in AGA for the
four examples presented in this work.}

\begin{tabular}{|c|c|c|c|c|}
\hline
\backslashbox{Example}{AGA's parameter}&
p&
$N_{1}$&
$N_{2}$&
$N_{0}$\tabularnewline
\hline
\multirow{2}{*}{Bi-dimensional functions (2)}&
\multirow{2}{*}{0.6}&
10&
9&
100\tabularnewline
&
&
20&
19&
400\tabularnewline
\hline
Extrasolar planets&
\multirow{2}{*}{0.45}&
\multirow{2}{*}{30}&
\multirow{2}{*}{29}&
\multirow{2}{*}{900}\tabularnewline
YSO&
&
&
&
\tabularnewline
\hline
\end{tabular}

\end{table}

In the first panel, we start with a population of $N_{0}=100$ individuals,
i. e., a set formed by 100 random points representing the candidate
solutions to the global maximum distributed more or less uniformly
in parameter space. After 5 generations (second panel), a clustering
at the second, third and fourth maxima is clearly apparent. After
10 generations, the solutions already cluster around the main maxima.
At the 25th generation, we have reached the maximum in $\left(x^{*},y^{*}\right)=\left(0.5,0.5\right)$
with $f\left(x^{*},y^{*}\right)=1$ with an accuracy of $\sim10^{-7}$.
We note that at the 25th generation all the 100 individuals have reached
the maximum with at least this accuracy.

In the last panel, we show the evolution in the fittest \emph{{}``phenotype''}
with the number of generations as plotted for two sizes of the population,
$N_{0}=100$ and $N_{0}=400$; in other words, we measure the deviation
in the function value for the maxima points identified in each generation
from the {}``true'' maximum, that is, $f\left(0.5^{*},0.5^{*}\right)=1$.
It is evident from Fig. \ref{fig:2} that a larger size of the population
causes the maximum to be reached in a lower number of generations.

\begin{figure*}
\centering

\includegraphics[width=15cm,height=15cm,keepaspectratio]{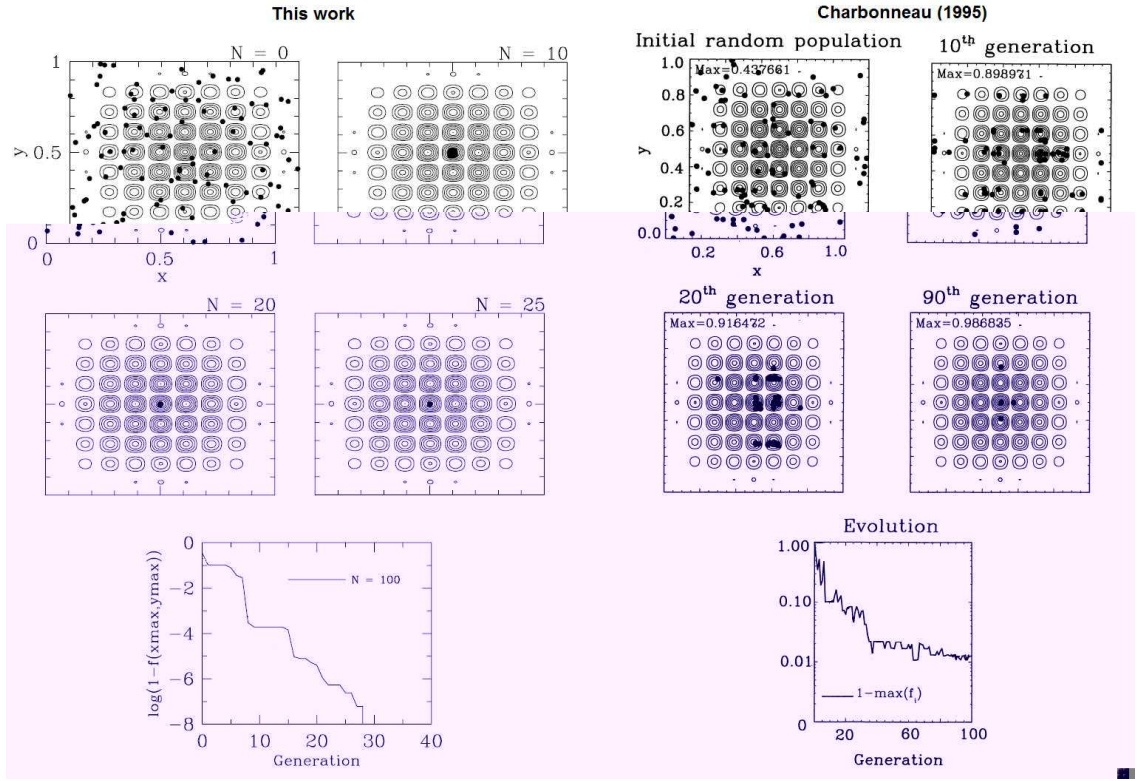}

\caption{Comparison of our results with those obtained by Charbonneau (\cite{Charbonneau}).
We have maintained the same format and removed the curve of the evolution
of the median-fitness individual in the fifth panel of Charbonneau's
work to facilitate direct comparison. In the first panel we start
with an initial population of 100 individuals and both algorithms
start to search the global maxima. While the GA employs 90 generations
to reach the solution, our AGA just requires 25 (fourth panels). The
fifth panels show the evolution of the fittest phenotype with the
number of generations for AGA (left) and GA (right). Note that indeed,
AGA reaches an accuracy of $\sim10^{-8}$ in less that 30 generations
while GA attains an accuracy of $\sim10^{-2}$ in 100 generations.}

\label{fig:3}
\end{figure*}

In Fig. \ref{fig:3}, we show the evolution in the global solution
using AGA and the results obtained by Charbonneau (\cite{Charbonneau})
using a traditional GA. In the first panel for both works, an initial
population of 100 individuals is randomly selected. After 10 generations,
all the individuals in our algorithm have clustered around the maxima;
in contrast, the GA has started to form some clusters. In the third
panel, after 20 generations, AGA has found the global maxima, while
the GA continues to search among clusters for a global solution. In
the fourth panel, the convergence to the global solution is shown
for both works, our AGA having reached the solution in 25 generations
with higher accuracy than the 90 generations employed by the GA. In
the fifth panel, we finally can compare the evolution in the fittest
phenotype for each generation, calculated to be $1-f\left(x_{max},y_{max}\right)$.
For AGA, these differences are much smaller than those obtained with
the GA method, which indicates that our algorithm is more accurate
than GA. However, we note that the solution presented in Charbonneau's
work was limited by the number of digits used in the encoding scheme.
The GA algorithm can also reach high levels of precision by changing
the encoding, but at the cost of slower convergence, which would make
the GA algorithm become even more computationally expensive than the
AGA. The lower number of generations required by AGA to reach a global
solution demonstrates that AGA is more efficient than the GA method.

\subsubsection{Example 2}

\begin{figure*}
\centering

\includegraphics[width=12cm,height=12cm,keepaspectratio]{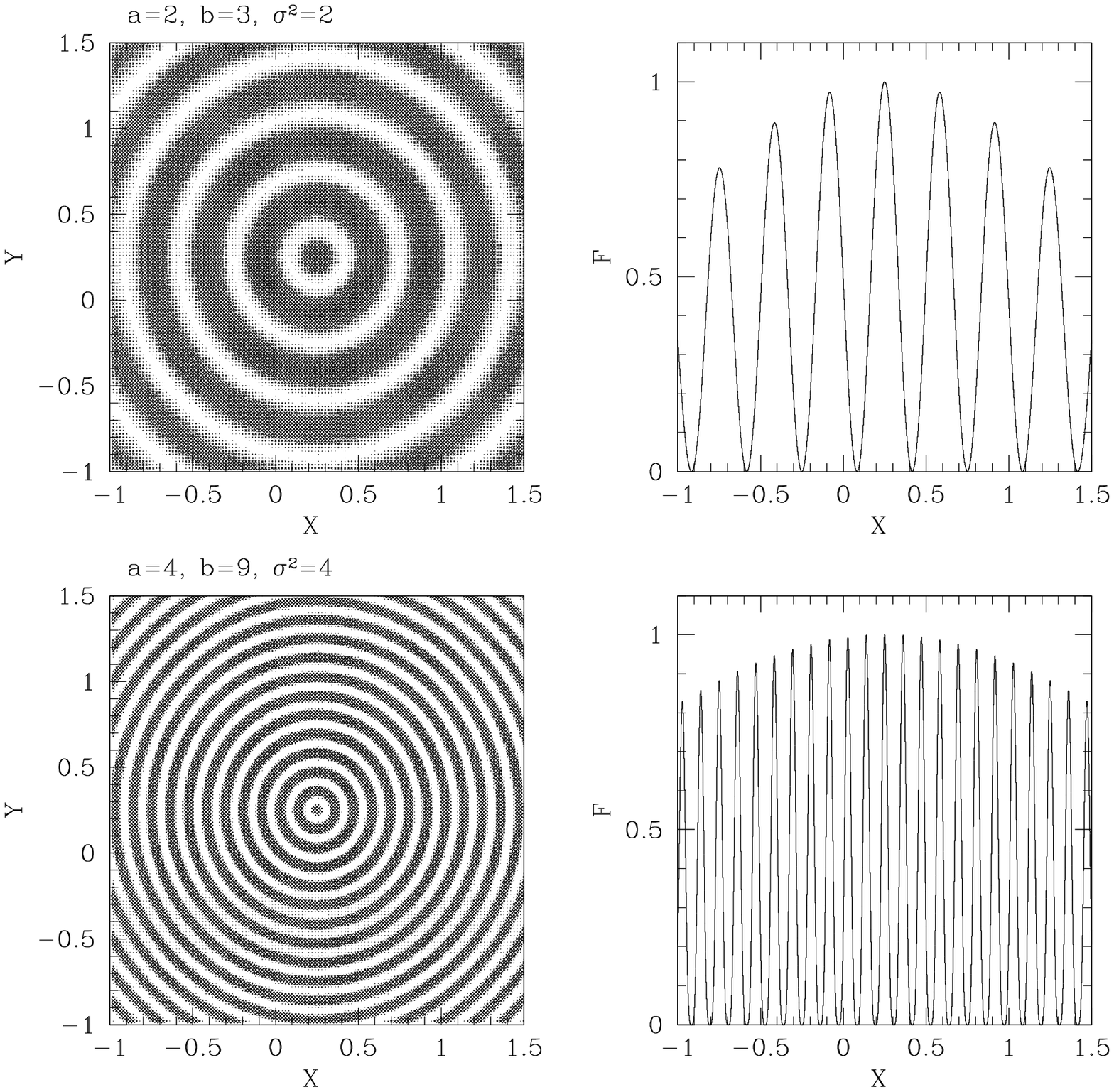}

\caption{Profile of the bi-dimensional positive cosine function. In the upper
panels, we take $a=2,\, b=3$ and $\sigma^{2}=2$; in the lower panels,
we take $a=4,\, b=9$ and $\sigma^{2}=4$. In both cases, the position
of the maximum peak is $\left(x,\, y\right)=\left(0.25,\,0.25\right)$,
where $f\left(0.25,\,0.25\right)=1$. }

\label{fig:4}
\end{figure*}

The following example is proposed as a function test in the PIKAIA's
user guide (Charbonneau and Knapp, \cite{Charbonneau and Knapp}).
The problem consists of locating the global maximum of the function:

\begin{equation}
f\left(r\right)=f\left(x,\, y\right)=cos^{a}\left(b\pi r\right)exp\left(-\frac{r^{2}}{2\sigma^{2}}\right)\label{eq:4}\end{equation}

\noindent where \textit{a} and \textit{b} are known constants, \textit{r}
is the radial distance given by the expression $r=\sqrt{\left(x-x_{0}\right)^{2}+\left(y-y_{0}\right)^{2}}$,
$\sigma$ is the width of the Gaussian and the position of the maximum
peak is $\left(x_{0},\, y_{0}\right)$. Observed from above, this
function appears like concentric rings of similar widths and amplitudes
(see Fig.\ref{fig:4}). The difficulty in finding the maximum peak
of this function is that the area of the concentric rings is much
larger than the area at the center containing the maximum peak (see
Fig.\ref{fig:4}). This property ensures that most of the algorithms
searching for the maximum peak fail because they usually become fixed
inside in one of the rings and identify a local maximum instead. We
experimented with different values of \emph{a}, \emph{b,} \textsl{\emph{and}}
$\sigma^{2}$ ($a=2,\,4$; $b=3,\,9$ and $\sigma^{2}=1,\,2,\,4$),
as well as different centers $\left(x_{0},\, y_{0}\right)$. In all
cases, using AGA with the parameters shown in Table \ref{tab:1},
the \textbf{}{}``true'' maximum peak was found in just a few steps,
typically in less than 100 generations, and with an accuracy of $10^{-6}$.
The results are summarized in Table \ref{tab:2}. Changing the position
of the center did not substantially change the number of generations
needed to reach the selected tolerance level.

\begin{table}
\centering

\caption{\label{tab:2} Values of the parameters \textit{a, b,} and $\sigma^{2}$
assumed for the function $f\left(r\right)=cos^{a}\left(b\pi r\right)exp\left(-\frac{r^{2}}{2\sigma^{2}}\right)$
and the number of generations needed to reach an error tolerance of
$10^{-6}$. }

\begin{tabular}{|c|c|c|c|}
\hline
\textit{a}&
\textit{b}&
$\sigma^{2}$&
Ngen\tabularnewline
\hline
\hline
2&
3&
1&
65\tabularnewline
\hline
2&
3&
2&
69\tabularnewline
\hline
2&
3&
4&
66\tabularnewline
\hline
2&
9&
1&
94\tabularnewline
\hline
2&
9&
2&
94\tabularnewline
\hline
2&
9&
4&
91\tabularnewline
\hline
4&
3&
1&
69\tabularnewline
\hline
4&
3&
2&
66\tabularnewline
\hline
4&
3&
4&
58\tabularnewline
\hline
4&
9&
1&
94\tabularnewline
\hline
4&
9&
2&
101\tabularnewline
\hline
4&
9&
4&
98\tabularnewline
\hline
\end{tabular}
\end{table}

\subsection{The parameter estimation problem in astronomical models}

In the physical sciences, curve or model fitting is essentially an
optimization problem. Giving a discrete set of $N$ data points $\left(x_{i},y_{i}\right)$
with associated measurement errors $\sigma_{i}$, one seeks the best
possible model (in other words, the closest fit) for these data using
a specific form of the fitting function, $y\left(x\right)$. This
function has, in general, several adjustable parameters, whose values
are obtained by minimizing a {}``merit function'', which measures
the agreement between the data and the model function $y\left(x\right)$.

We suppose that each data point $y_{i}$ has a measurement error that
is independently random and distributed as a normal distribution about
the {}``true'' model with standard deviation $\sigma_{i}$. The
maximum likelihood estimate of the model parameters $\left(c_{1},...,c_{k}\right)$
is then obtained by minimizing the function,

\begin{equation}
\chi^{2}\equiv{\displaystyle \underset{i=1}{\overset{N}{\sum}}\left(\frac{y_{i}-y\left(x_{i};\, c_{1},..,c_{k}\right)}{\sigma_{i}}\right)^{2}}\label{eq:5}\end{equation}

\subsubsection{Evaluation of the error estimation}

The experimental or observational data are subject to measurement
error, thus it is desirable to estimate the parameters in the chosen
model and their errors. In the straight-line-data fitting and the
general linear least squares,we can compute the standard deviations
or variances of individual parameters through simple analytic formulae
(Press et al., \cite{Press}). However, when we attempt to minimize
a function such as Eq. \ref{eq:5}, we have no expression for calculating
the error in each parameter. A good approach to solve this problem
consists of building {}``synthetic data sets''. The procedure is
to draw random numbers from appropriate distributions so as to mimic
our clearest understanding of the underlying process and measurement
errors in our apparatus. With these random selections, we compile
data sets with exactly the same numbers of measured points and precisely
the same values of all control or independent variables, as our true
data set. In other words, when the experiment or observation cannot
be repeated, we simulate the results we have obtained.

\begin{figure}
\centering

\includegraphics[width=8cm,height=8cm,keepaspectratio]{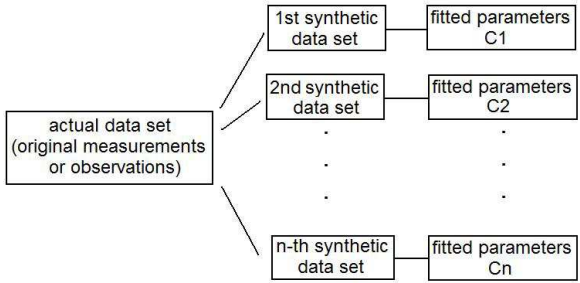}

\caption{From the original set of measurements/observations, several synthetic
data sets are constructed. This is achieved by adding to the dependant
variable a random number whose absolute value is within the estimated
errors of the original data. For each of these synthetic data sets,
a set of fitted parameters is obtained. The average value of each
parameter and the corresponding standard deviation are taken as estimates
of the parameters and their associated errors.}

\label{fig:5}
\end{figure}

We compiled the synthetic data with the process illustrated in Fig.
\textbf{\ref{fig:5}} and from the following expression\textbf{:}

\begin{equation}
y_{i}^{'}=y_{i}+\sigma_{i}\left(2\xi-1\right)\label{eq:6}\end{equation}

\noindent where $y_{i}^{'}$ represents the i-th data point of the
new set, $y_{i}$ is the i-th data point of the original set, $\sigma_{i}$
is the error associated with the i-th data point, and $\xi$ is a
random number within the interval $\left[0,1\right]$. Using Eq. \ref{eq:6},
we generated synthetic data sets with the original errors because
they are related to the measurements.

\noindent For each of these data sets, we found a corresponding set
of parameters (see Fig. \ref{fig:5}). Finally, we calculated the
average values of each parameter and the corresponding standard deviation,
as the estimates of the parameters and their associated errors, respectively.
We applied this procedure to the following examples.

\subsubsection{Fitting the orbits of Extrasolar Giant Planets}

We use the merit function given in Eq. \ref{eq:5} and the algorithm
described in Sect. 2 to solve an interesting and challenging task
in astronomy: curve or model fitting of data for the orbits of the
extrasolar planets.

The first extrasolar planet was discovered in 1995 by Mayor and Queloz
(\cite{Mayor}) and, according to The Extrasolar Planets Encyclopaedia%
\footnote{http://exoplanet.eu/catalog%
}, until February 2009 there had been 342 candidates detected. Most
of them (316) were revealed by radial velocity or astrometry of 269
host stars, that is, by Keplerian Doppler shifts in their host stars.
Doppler detectability favors high masses and small orbits depending
mainly on the present Doppler errors achievable with available instruments.
In addition, the precision of the Doppler technique is probably about
3 m s$^{\text{-1}}$, owing to the intrinsic stability limit of stellar
photospheres. This technique is sensitive to companions that induce
reflex stellar velocities, $K>10\,\mathrm{m\, s^{-1}}$, and exhibit
orbital periods ranging from a few days to several years, the maximum
detectable orbital period being set by the time baseline of the Doppler
observations. The remaining exoplanets were detected by other techniques:
microlensing (8), imaging (11), and timing (7). For this example we
only refer to the planets detected by radial velocity.

We now consider a system consisting of a central star of mass $M_{*}$,
surrounded by $N_{p}$ planets in bounded orbits. Assuming that the
orbits are unperturbed Kepler orbits, the line-of-sight velocity of
the star relative to the observer (Lee and Peale, \cite{Lee and Peale})
is,

\begin{equation}
v_{*}=v_{0}+\sum_{j=1}^{N_{p}}K_{j}\left[cos\left(f_{j}+\omega_{j}\right)+e_{j}cos\,\omega_{j}\right]\label{eq:7}\end{equation}

\noindent where $e_{j}$ is the eccentricity, $\omega_{j}$ is the
argument of the pericenter, $f_{j}$ is the true anomaly, $K_{j}$
is the velocity amplitude of planet \emph{j}, and $v_{0}$ is the
line-of-sight velocity of the center-of-mass relative to the observer.
The true anomaly is related to time by means of Kepler's equation,

\begin{equation}
E-e_{j}\, sin\left(E\right)=\frac{2\pi}{T_{j}}\left(t-t_{0j}\right)\label{eq:8}\end{equation}

\noindent and

\begin{equation}
tan\left(\frac{f_{j}}{2}\right)=\sqrt{\frac{1+e_{j}}{1-e_{j}}}tan\left(\frac{E}{2}\right)\label{eq:9}\end{equation}

\noindent where $T_{j}$ and $t_{0j}$ are the period and time, respectively,
of pericenter passage for planet \emph{j.} Thus, for each planet,
there are \emph{5 free parameters}: $e_{j}$ , $T_{j}$, $t_{0j}$,
$K_{j}$ and $\omega_{j}$. Additionally, there is the systemic velocity
$v_{0}$. In total, we have $\left(5\times N_{p}\right)+1$ free parameters,
which are simultaneously fitted using AGA.

When these basic parameters for each planet are known through model
fitting, we can estimate the semi-major axes of the orbits, $a_{j}$,
and the masses of each planet, $m_{j}$. To do this, we have to make
other simplifying assumptions. For instance, in the simplest model
of totally independent planets, \begin{equation}
K_{j}=\left(\frac{2\pi G}{T_{j}}\right)^{\frac{1}{3}}\frac{m_{j}sin\left(i_{j}\right)}{\left(M_{*}+m_{j}\right)^{\frac{2}{3}}}\frac{1}{\sqrt{1-e_{j}^{2}}}\label{eq:10}\end{equation}

\noindent from which we can determine either $m_{j}$ by assuming
$sin\, i_{j}=1$ or\, $m_{j}sin\, i_{j}$ by neglecting $m_{j}$
in front of the mass of the star, $M_{*}$. We can then estimate $a_{j}$
using Kepler's third law,

\begin{equation}
T_{j}^{2}=\frac{4\pi^{2}a_{j}^{3}}{G\left(M_{*}+m_{j}\right)}\label{eq:11}\end{equation}

\noindent where \emph{G} denotes the gravitational constant.

We applied the AGA presented in Sect. 2 to the measured radial velocities
of the main-sequence star 55 Cancri, published by Fischer et al. (\cite{Fischer & Marcy}).

\begin{figure*}
\centering

\includegraphics[width=11cm,height=11cm,keepaspectratio]{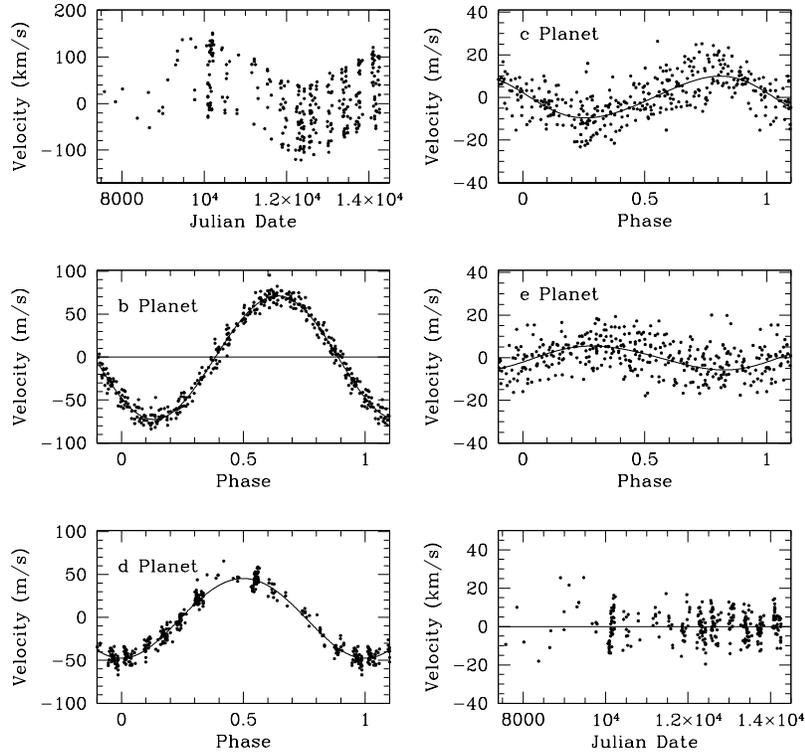}

\caption{The velocities and fits for each of the four planets are shown separately
for clarity by subtracting the effects of the other planets. That
is, for planet labelled \textit{b}, planets \textit{c}, \textit{d}
and \textit{e} have been removed from the data using the parameters
found in the simultaneous four-planet fitting. For planet labelled
\textit{c} we have removed planets \textit{b}, \textit{d} and \textit{e.}
For planet labelled \textit{d} we have removed planets \textit{b},
\textit{c} and \textit{e}. For planet labelled \textit{e} we have
removed planets \textit{b}, \textit{c} and \textit{d}. We have added
the fitted systemic velocity for the star, which value is 17.2826
m s$^{\text{-1}}$. The last panel (bottom and right) shows the residuals
between the data and the model. The first (up and left) panel shows
the raw data.}

\label{fig:6}
\end{figure*}

The data set contains 250 measurements completed at the Lick Observatory
from 1989 to 2007, and 70 measurements made at the Keck Observatory
from 2002 to 2007 of the velocity of 55 Cancri. Data from Lick has
measurement errors of $\sim10\,\mathrm{m\, s^{-1}}$ (1989-1994) and
3-5 $\mathrm{m\, s^{-1}}$ (1995-2007). Data from Keck has measurement
errors of 3 $\mathrm{m\, s^{-1}}$ for data acquired prior to 2004
August and 1.0-1.5 $\mathrm{m\, s^{-1}}$ thereafter. The measured
data have a maximum amplitude of $\sim150\,\mathrm{m\, s^{-1}}$ and
are of excellent signal-to-noise ratio. The orbital parameters were
established by the detection of four planets in all Doppler measurements.
To test a possible stellar systemic velocity remnant in the data,
we added an additional parameter to the systemic velocity of the star
obtaining a value of 17.28 $\mathrm{m\, s^{-1}}$. We fitted the orbits
of 1, 2, 3, and 4 planets to this data (see Fig. \ref{fig:6}). This
exercise shows that the fitting solution improves when considering
the orbit of more planets. The rms of the residuals and the $\chi_{red}$,
$\left(\chi_{red}^{2}=\frac{\chi^{2}}{N_{points}-N_{parameters}-1}\right)$
values, improve from values of 33.5 and 9.72, respectively, when fitting
the orbit of one planet, to values of 7.99 and 2.03, respectively,
when fitting the orbit of four planets (see Table \ref{tab:3}). In
general, the values of the rms, $\sqrt{\chi_{red}^{2}}$ fits and
the derived parameters compare well with those obtained by Fischer
et. al. (\cite{Fischer & Marcy}). In Tables \ref{tab:3} and \ref{tab:4},
we summarize our results.

\begin{table}
\centering

\caption{\label{tab:3}Our values of the rms of the residuals and the $\sqrt{\chi_{red}^{2}}$
fits for the orbital fitting problem using AGA. For comparison, in
the last two columns we show their corresponding values obtained by
Fischer et al. (\cite{Fischer & Marcy}).}

\begin{tabular}{ccccc}
\hline
\multirow{2}{*}{Planet}$^{\text{1}}$&
This work&
\multicolumn{1}{c}{This work}&
\multicolumn{1}{c}{Fischer et al.}&
\multicolumn{1}{c}{Fischer et al.}\tabularnewline
&
rms (m s$^{\text{-1}}$) &
$\sqrt{\chi_{red}^{2}}$ &
rms (m s$^{\text{-1}}$)&
$\sqrt{\chi_{red}^{2}}$\tabularnewline
\hline
b&
33.5&
9.72&
39&
10\tabularnewline
b, c&
10.69&
3.38&
11.28&
3.42\tabularnewline
b, c, d &
9.69&
2.49&
8.62&
2.50\tabularnewline
b, c, d, e&
7.99&
2.03&
7.87&
2.12\tabularnewline
\hline
\end{tabular}

$^{\text{1}}$The single-planet fitting was obtained in Marcy et al.
(\cite{Marcy}).
\end{table}

With the exceptions of the third and fourth planet fittings, our rms
values are lower than those obtained by Fischer et al. (\cite{Fischer & Marcy}).
Our $\sqrt{\chi_{red}^{2}}$ values are however lower in all cases.

\begin{table*}
\centering

\caption{\label{tab:4}Our estimated values for the five parameters of the
four exoplanets around 55 Cancri. The planets are listed in order
of increasing orbital period, and the planet designations, b-e, correspond
to the notation given by Marcy et al. (\cite{Marcy}) and Fischer
et al. (\cite{Fischer & Marcy}). The value for the $\sqrt{\chi_{red}^{2}}$
is also included. \protect \\
}

\begin{tabular}{ccccccc}
\hline
Planet&
$T$

(days)&
$t_{0}$

(JD)&
$e$ &
$\omega$

(deg)&
$K_{1}$

(m s$^{\text{-1}}$)&
$\sqrt{\chi_{red}^{2}}$\tabularnewline
\hline
e&
2.8170 $\pm$ 0.0932&
10000.0046 $\pm$ 0.3250&
0.07 $\pm$ 0.0016&
250.2326 $\pm$ 0.4613&
5.4311 $\pm$ 0.2241&
\multirow{4}{*}{2.0314}\tabularnewline
b&
14.6515 $\pm$ 0.0002&
10002.8917 $\pm$ 0.0549&
0.0145 $\pm$ 0.0005&
130.9176 $\pm$ 0.4591&
71.7606 $\pm$ 0.3140&
\tabularnewline
c&
44.3298 $\pm$ 0.0059&
9989.9237 $\pm$ 0.4637&
0.0853 $\pm$ 0.0014&
78.2384 $\pm$ 0.4662&
9.9820 $\pm$ 0.2075&
\tabularnewline
d&
5218.3339 $\pm$ 0.5246&
12500.7572 $\pm$ 0.3713&
0.0250 $\pm$ 0.0006&
180.2123 $\pm$ 0.2645&
46.6872 $\pm$ 0.1431&
\tabularnewline
\hline
\end{tabular}
\end{table*}

From the parameters shown in Table \ref{tab:4} we were able to derive
the mass of the planet, $M_{P}sin\, i$ (M$_{\text{J}}$), and the
major semiaxis \textit{a} (AU). For the four-planet fitting, we found
(Table \ref{tab:5}) that our values compare well with those reported
by Fischer et al. (\cite{Fischer & Marcy}) in their five-planet model.
Except for the first planet, all the values for the major semiaxis
have a standard error lower than those obtained in the Fischer's model.
In the case of the mass of each planet, our values have smaller standard
errors.

\begin{figure}
\centering

\includegraphics[width=8cm,height=10cm,keepaspectratio]{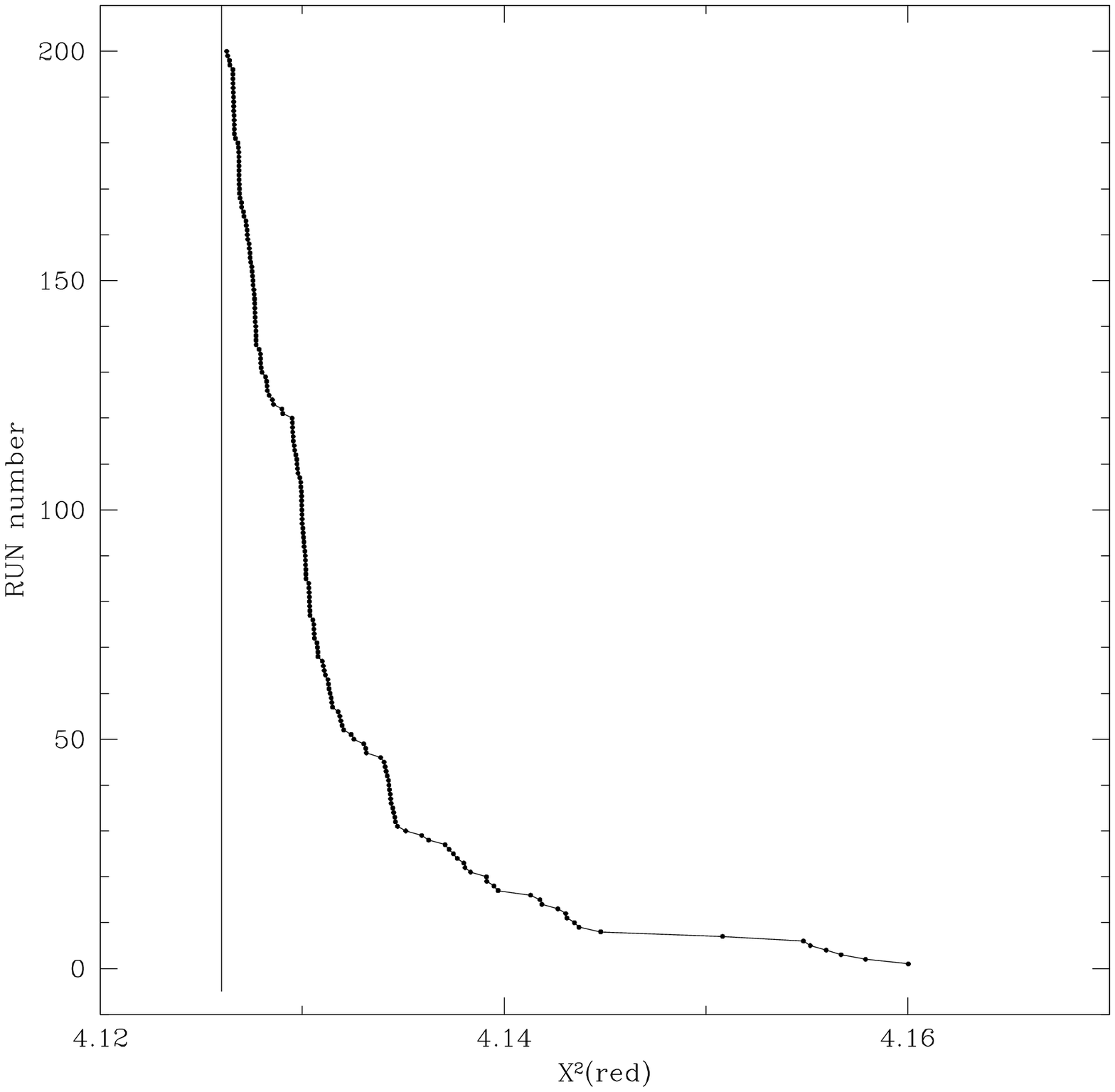}

\caption{Convergence of the $\chi_{red}^{2}$ value with the number of runs
(do not confuse with the number of generations done within AGA). The
curve shows how $\chi_{red}^{2}$ converges to a {}``limiting''
value in the sense that the variations occur in the last decimals
(vertical line). As the number of runs increases the value of $\chi_{red}^{2}$
diminishes which means that we have not reached the optimal solution.
Only after about 100 runs the $\chi_{red}^{2}$ value has converged
and at this point the parameters do not significally change after
subsecuent runs.}

\label{fig:7}
\end{figure}

We included the iterative scheme discussed in sect. 2 in fitting the
orbit of the planets. In Fig. \ref{fig:7}, we show the value of the
$\chi_{red}^{2}$ for the four-planet fit calculated after each run.
In this case, we started with $\chi_{red}^{2}=4.16$ and after the
10th iterative run, the $\chi_{red}^{2}$ had diminished to 4.15,
which means that we had not found the optimal solution. During the
first iterative runs, we note that the value of $\chi_{red}^{2}$
decreased rapidly. After about 100 iterative runs, the $\chi_{red}^{2}$
had not changed significally and converged to a fixed value (there
are only slight variations in the last decimals). At this point, we
can be assured that the value of the parameters have been found. Finally,
we estimated the errors in fitting the orbits of the four planets
as described in sect. 3.2.1. We present the estimated errors in Tables
\ref{tab:4} and \ref{tab:5}.\\

\begin{table*}
\centering

\caption{\label{tab:5}Our derived values for the mass of the planet and the
major semiaxis for the four-planet fitting. For comparison, we also
show the values obtained by Fischer et al. (\cite{Fischer & Marcy})
for their five-planet model, where we have removed the fifth body.
\protect \\
}

\begin{tabular}{ccccc}
\hline
\multirow{2}{*}{Planet}&
This work&
This work&
Fischer et al. &
Fischer et al. \tabularnewline
&
$M_{P}sin\, i$ (M$_{\text{J}}$)&
\textit{a} (AU)&
$M_{P}sin\, i$ (M$_{\text{J}}$)&
\textit{a} (AU)\tabularnewline
\hline
e&
0.0361$\pm$0.0014&
0.0383$\pm$0.0008&
0.034$\pm$0.0036&
0.038$\pm$$1.0\times10^{-6}$\tabularnewline
b&
0.8285$\pm$0.0036&
0.1148$\pm$$1.0\times10^{-6}$&
0.824$\pm$0.007&
0.115$\pm$$1.1\times10^{-6}$\tabularnewline
c&
0.1661$\pm$0.0035&
0.2402$\pm$$2.1\times10^{-5}$&
0.169$\pm$0.008&
0.240$\pm$$4.5\times10^{-5}$\tabularnewline
d&
3.8201$\pm$0.0117&
5.7705$\pm$0.0004&
3.835$\pm$0.08&
5.77$\pm$0.11\tabularnewline
\hline
\end{tabular}
\end{table*}

\subsubsection{Fitting the spectral energy distribution (SED) for a YSO}

A spectral energy distribution (SED) is a plot of flux, brightness
or flux density versus frequency/wavelength of light. It is widely
used to characterize astronomical sources i. e., to identify the type
of source (star, galaxy, circumstellar disk) that produces these fluxes
or brightness. Modelling the observed SEDs can help us to infer the
temperature and size, among the other physical parameters of the source.
As examples in radio astronomy, a SED with a negative spectral index
$\sim-0.7$ would indicate the presence of a synchrotron radiation
source; in infrared astronomy, SEDs can be used to classify T-Tauri
stars; in galactic astronomy, the analysis of the SEDs leads to the
determination of the respective roles of the old and young stellar
populations in dust-grain heating.

For the reasons explained above, it is interesting to find the adequate
model for an observed SED. We consider the observations reported by
Curiel et al. (\cite{Curiel09}) in the L1448 region. This cloud is
part of the Perseus molecular cloud complex located at a distance
of $\sim\mathrm{300\, pc}$. We are interested in fitting the observed
SED for a couple of reasons: it contains an extremely young and highly
collimated bipolar outflow (Bachiller et al. \cite{Bachiller}) and
seems to be a site of very recent star formation based on some observations
(Anglada et al. \cite{Anglada}; Curiel et al. \cite{Curiel}; Barsony
et al. \cite{Barsony}; Girart \& Acord \cite{Girart and Acord};
Reipurth et al. \cite{Reipurth}).

The data are taken from Curiel et al. (\cite{Curiel09}), who carried
out a fit of the SED by assuming that there are three main components
contributing to the flux at different wavelengths. Thus, the fitting
function was given by the contribution of an optically free-free component
and two grey bodies:

\begin{equation}
S_{\nu}=c_{1}\omega^{c_{2}}+\frac{c_{3}\left(1-e^{-c_{4}\omega^{c_{5}}}\right)\omega^{3}}{e^{c_{6}\omega}-1}+\frac{c_{7}\left(1-e^{-c_{8}\omega^{c_{9}}}\right)\omega^{3}}{e^{c_{10}\omega}-1}\label{eq:12}\end{equation}

\noindent where $\omega\equiv\frac{\nu}{\nu_{0}}$ is the frequency
normalized to a reference frequency, $\nu_{0}$.

\noindent The free parameters to be estimated, by the minimization
of the chi-square (Eq. \ref{eq:5}), are identified as $c_{1},c_{2},...,c{}_{10}$.
They are defined in terms of their corresponding physical parameters
as: $c_{1}\equiv F_{C}$ is the flux of optically thin emission with
a spectral index given by $c_{2}\equiv\alpha$, which corresponds
to the free-free emission coming from a thermal jet;\,$c_{3}=F_{1}\equiv\frac{2h\Omega\nu_{0}^{3}}{c^{2}}$
is a reference flux;\, $c_{4}\equiv\tau_{1}$ is the dust opacity
evaluated at the reference frequency $\nu_{0}$;\,$c_{5}\equiv\beta_{1}$
is the dust emissivity index; and $c_{6}\equiv\frac{h\nu_{0}}{kT_{1}}$
is related to the temperature. The last \textbf{}four parameters correspond
to the emission originating in the first grey body (probably a molecular
envelope). Finally, we propose that a second grey body exists that
is a circumstellar disk surrounding the young protostar. Similarly,
the parameters $c_{7}$, $c_{8}$, $c_{9}$, and $c_{10}$ are related
to the physical parameters of the second grey body.

\begin{figure}
\centering

\includegraphics[width=8cm,height=10cm,keepaspectratio]{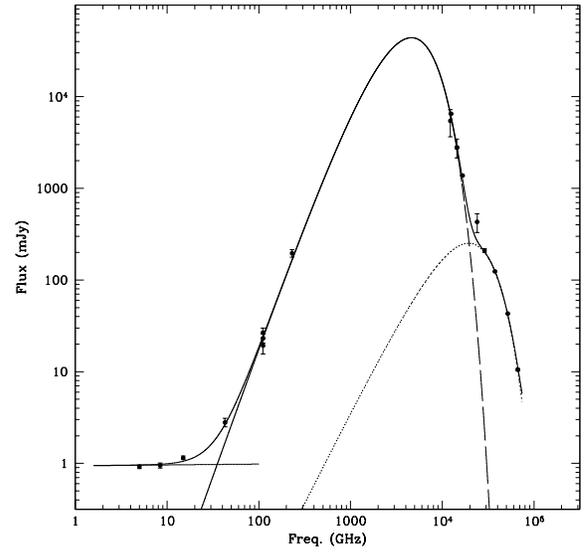}

\caption{The Spectral Energy Distribution (SED) for a YSO in the L1448 region.
The dots represent the observations and the bars their associated
measurement errors. The straight line corresponds to the continuum
flux, the dashed line is the first grey body, the dotted line is the
second grey body and the solid line is the sum of the three contributions.}

\label{fig.8}
\end{figure}

\noindent In the previous expressions, $\nu$ is the frequency at
which the source is observed, $h$ is the Planck's constant, $\Omega$
is the solid angle subtended by the source, $c$ is the speed of light,
$\tau$ is the optical depth at the frequency $\nu_{0}$ for each
component, $\beta$ is the dust emissivity index for each component,
and $k$ is the Boltzmann's constant. We assumed the characteristic
frequency of the source to be $\nu_{0}=\frac{c}{\lambda_{7mm}}$ (Curiel
et. al., \cite{Curiel09}). Table \ref{tab:6} summarizes the values
of the fitted parameters in the model described by Eq. \ref{eq:12}
of the SED for a YSO in the L1448 region. Table \ref{tab:6} also
includes the estimated errors in the fitted parameters, as described
in sect. 3.3.1. Figure \ref{fig.8} shows the observed data together
with the best fit.

\begin{table}
\centering

\caption{\label{tab:6} Estimated values for the ten parameters in the model
of the SED for a YSO in the L1448 region.}

\begin{tabular}{ccc}
\hline
Parameter&
Value&
$\sqrt{\chi_{red}^{2}}$\tabularnewline
\hline
$F_{C}$ (mJy)&
0.9295 $\pm$ 0.0330&
\multirow{10}{*}{1.9720}\tabularnewline
$\alpha$&
-0.0169 $\pm$ 0.0166&
\tabularnewline
$F_{1}$ (mJy)&
1.7109$\pm$ 0.1141&
\tabularnewline
$\tau_{1}$&
0.0728$\pm$ 0.0062&
\tabularnewline
$\beta_{1}$&
0.8999 $\pm$ 0.0575&
\tabularnewline
$T_{dust\,1}$ (K)&
75.7522 $\pm$ 0.5273&
\tabularnewline
$F_{2}$ (mJy)&
0.0068 $\pm$ 0.0008&
\tabularnewline
$\tau_{2}$&
34.9140 $\pm$ 40.7209&
\tabularnewline
$\beta_{2}$&
1.5992 $\pm$ 0.3254&
\tabularnewline
$T_{dust\,2}$ (K)&
330.8240 $\pm$ 1.2946&
\tabularnewline
\hline
\end{tabular}
\end{table}

From Fig. \ref{fig.8}, we can conclude that the fitted model with
the parameters shown in Table \ref{tab:6} is adequate for explaining
the observations of the YSO in the L1448 region. The fitted curve
lies within the measurement error bars (except one point located at
high frequencies) and the SED consists of a continuum component and
the emission from two grey bodies. Further astrophysical implications
of these results can be found in Curiel et al. (\cite{Curiel09}).

\section{Summary and Conclusions}

We have presented a simple algorithm based on the idea of genetic
algorithms to optimize functions. Our algorithm differs from standard
genetic algorithms mainly in mainly two respects: 1) we do not encode
the initial information (that is, the initial set of possible solutions
to the optimization problem) into a string of binary numbers and 2)
we propose an asexual reproduction as a means of obtaining new {}``individuals''
(or candidate solutions) for each generation.

We have then applied the algorithm in solving two types of optimization
problems: 1) finding the global maximum in functions of two variables,
where the typical techniques fail, and 2) parameter estimation in
astronomy by the minimization of the chi-square. For the latter case,
we considered two examples: fitting the orbits of extrasolar planets
associated with the star 55 Cancri and fitting the SED of a YSO for
the L1448 region.

We found that our algorithm has several advantages:

\begin{itemize}
\item It is easy to implement in any computer because it does not require
an encoding/decoding routine, and the new generations are constructed
by the asexual reproduction of a selected subset (with the highest
fitness) of the previous population.\\

\item The algorithm does not require the evaluation of standard genetic
operations such as crossover and mutation. This is replaced by a set
of sampling rules, which simplifies the creation of new generations
and speeds up the finding of the best solution. \\

\item When the initial {}``guess'' is far from the solution, AGA is capable
to migrate searching for the {}``true'' optimal solution. The final
solution is usually achieved after a few hundred generations, in some
cases, even faster.\\

\item In some difficult cases, such as the fitting of the orbits of several
exoplanets and the fitting of the SED of YSOs (with several components),
AGA finds the solution in a few hundred generations. In these cases,
an iterative scheme (several runs of AGA using the solutions of one
run as the initial {}``guess'' of the next run) can help to improve
finding the {}``true'' solution. This is particularly useful when
the variables have different orders of magnitude, which causes the
different variables to not converge to the final solution at the same
time.\\

\item As a consequence of the previous points, AGA becomes computationally
less expensive than the standard version (GA). The convergence of
AGA is reached in just a few generations.
\end{itemize}
From the example of the orbital fitting for exoplanets around 55 Cancri,
we can conclude that our algorithm gives parameters that compare well
with those obtained by Fischer et al. (\cite{Fischer & Marcy}). Marcy
et al. (\cite{Marcy}) also used the standard GA to obtain the parameters
for a third planet, and they concluded that the GA fails to significally
improve the $\chi^{2}$ value.

In the case of the SED for a YSO in the L1448 region, we find that
the most adequate model contains the contribution of two grey bodies
at high frequencies and free-free continuum emission, through a power
law, at low frequencies.

Generally speaking the implementation of any kind of GA is advantageous
in the sense that we do not have to compute any derivative of the
selected fitness function allowing us to optimize functions with several
local maxima points.

In this work, we have also implemented a method to estimate the error
associated with each parameter based on the generation of {}``synthetic
data sets''. This method is easy to implement in any type of measured/observed
data and constitutes a way of estimating the errors in the parameters
when the minimization of the $\chi^{2}$ is employed.

\begin{acknowledgements}
We acknowledge the support of CONACyT (Mexico) grants 61547 and 60581.
EMG also thanks to DGAPA-UNAM postdoctoral fellowship for the financial
support provided for this research. We thank the anonymous referee
for the helpful comments made to this work.
\end{acknowledgements}

\end{document}